# Comments on the continuing widespread and unnecessary use of a defective emission equation in field emission related literature

Richard G. Forbes

Advanced Technology Institute & Department of Electrical and Electronic Engineering, University of Surrey, Guildford, Surrey, GU2 7XH, UK

r.forbes@trinity.cantab.net

**ABSTRACT**

Field electron emission (FE) has relevance in many different technological contexts. However, many related technological papers use a physically defective elementary FE equation for local emission current density (LECD). This equation takes the tunneling barrier as exactly triangular, as in the original FE theory of 90 years ago. More than 60 years ago, it was shown that the so-called Schottky-Nordheim (SN) barrier, which includes an image-potential-energy term (that models exchange-and-correlation effects) is better physics. For a metal-like emitter with work-function 4.5 eV, the SN-barrier-related Murphy-Good FE equation predicts LECD values that are higher than the elementary equation values by a large factor, often between around 250 and around 500. By failing to mention/apply this 60-year-old established science, or to inform readers of the large errors associated with the elementary equation, many papers (aided by defective reviewing) spread a new kind of "pathological science", and create a modern research-integrity problem. The present paper aims to enhance author and reviewer awareness by summarizing relevant aspects of FE theory, by explicitly identifying the misjudgment in the original 1928 Fowler-Nordheim paper, by explicitly calculating the size of the resulting error, and by showing in detail why most FE theoreticians regard the 1950s modifications as better physics. Suggestions are made, about nomenclature and about citation practice, that may help to diminish misunderstandings.

## I. INTRODUCTION

*Field electron emission* (FE)—sometimes called "cold FE" or often simply "field emission"—is a well-known electron emission process[1-4] in which most electrons escape by electrostatic-field-induced tunneling through a rounded triangular potential-energy (PE) barrier, at normal-energy levels significantly below the top of the tunneling barrier. This process is often known as "Fowler-Nordheim tunneling".

Field electron emission occurs in very many different technological contexts, either constructively, as a means of creating an electron source, or destructively, as a cause of electrical breakdown. In both cases, it would be helpful, either for design purposes or for breakdown-analysis purposes, to have equations that reliably describe expected FE current densities and currents.

However, the situation is confused by the widespread use in FE literature of a physically inadequate elementary emission equation [eq. (4) below]. This occurs most commonly in the literature of technological applications to which FE is or may be relevant. As compared with corrected theory, this elementary equation usually under-predicts local current densities by a large factor, which—for a metal-like emitter with work function 4.5 eV—would often be between around 250 and around 500. For emission out of a non-degenerate semiconductor conduction band—where electron affinity, rather than work-function, is the relevant parameter—a similar physical problem exists in principle, but the errors (not discussed in detail here) would be somewhat less,

The physically inadequate elementary equation is a simplified version of the original FE equation developed by Fowler and Nordheim[5,6] (FN) in 1928/29 [eq. (3) below], which is also physically inadequate. The problem is partly due to small misjudgments in physical thinking by FN (see Appendix A), whose analysis led to a mathematically correct result for an unphysical model potential-energy variation.

Calculations by Nordheim[7] later in 1928 used a more realistic PE model, but unfortunately contained a mathematical mistake that also led to under-prediction of local current densities.

The existence of these problems has been known[8] since 1953, and a corrected FE equation [eq. (5) below] was put in place by Murphy and Good[9] (MG) in 1956. However, 60 years later, the defective elementary FE equation is widely used in relevant technological literature, even though the (better) Murphy-Good FE equation is normally used in theoretical FE papers and in FE research textbooks.

More generally, there is growing concern over the integrity of scientific research (e.g., ref. 10 and references therein, and ref. 11). As part of this, it seems likely that there may develop concern over the reliability of scientific reviewing, and, later, political interest in this issue. It seems to the author that the present situation in FE can be seen as a mild theoretical variant of "pathological science" as described by Langmuir[12] and more recently by Park[13] and by Charlton[11], and that technological field emission has a research-integrity problem.

By "pathological science", I mean continuing situations where groups of scientists firmly convince themselves of the validity of some assumption or the adequacy of some approximation, when (a) there is strong independent scientific evidence that the assumption/approximation is not true/adequate, and (b) scientists outside the group do not accept the validity of the assumption/approximation. This is an extension of the term "pathological science" to be a neutral name that includes theory-oriented situations where community effects, as discussed in refs 11-13, seem to be operating (though I do not subscribe to the extreme viewpoint adopted by Charlton[11]). The more usual use of the term is in connection with misinterpreted experiments, by individuals or by groups. Langmuir and Hall[12] describe various examples, and the "cold fusion" episode (as competently analyzed by Park[13]), where numbers of scientists came to believe that the effect would eventually generate significant power, is a more recent example.

It can also be argued that journals and learned societies that uncritically publish papers containing inadequate equations, when it has been firmly established for 60 years that these equations are inadequate (and that much better equations are known), also have a research-integrity problem—and might find it difficult to justify their actions to a well-informed formal external enquiry.

Precisely why this use of a defective FE equation persists is extremely unclear, and probably there is no single reason. It may be that some researchers (both authors and reviewers) are simply unaware that the equation is inadequate, or are unaware of the size of the error, or regard omission of a factor of value 300 (or thereabouts) as an "approximation" not important in their particular research topic, or—when developing technology—regard it as better to use an equation that is simple rather than a more accurate one that is marginally more complicated. Or researchers may not accept the scientific validity of the 1953/56 corrections and simply ignore them. Or researchers may be unduly influenced by statements in the original 1928 FN paper (see Section IVD below). Alternatively, as suggested by Charlton[11], it may be that some researchers both (a) feel it best to follow community practice by working with what is written in the literature of the immediate subject area, *and* (b) do not consider it necessary (or do not have time) to conduct checks or report on whether the wider scientific community thinks their material is correct science.

Whatever the cause or causes, I regard it as vital that this use of a clearly defective FE equation be discontinued in research contexts, as soon as may be achieved. The aims of the present paper are to provide information intended to help this process and to help and encourage reviewers to challenge the use of defective FE equations in research contexts, by using the arguments set out here if it helps.

In what follows, the paper describes more of the general background relating to FE equations and their limitations, briefly summarizes some relevant developments in FE theory over the last 90 years, explicitly demonstrates the size of the error, explains why FE theoreticians (and other scientists outside the subject area) believe that the 1950s physical assumptions are correct (and hence that Murphy-Good theory is better physics than the original Fowler-Nordheim theory), and makes some suggestions about nomenclature and citation practice that may help to diminish confusion.

Appendix A discusses relevant weaknesses in the 1928 papers.

In accordance with common practice in electron emission literature, this paper uses the "electron emission convention" in which "field" $F$, "current density" $J$, and "current" $I$ are taken as positive quantities that are the negatives of the corresponding quantities as defined in conventional electrical theory. For reference, all universal constants used are given to 7 significant figures, but it is expected that these will be appropriately approximated in practical applications.

## II. THEORETICAL BACKGROUND

### A. General emission theory background

In many cases, real field emitters are needle-shaped or post-shaped, with rounded ends. The emitter apex radius of curvature affects tunneling probabilities, and also affects issues relating to emission area. However, past current-voltage data analysis has normally used the *planar emitter approximation*, which assumes that emitters can be modeled as having a smooth flat planar surface, with effectively constant emission area. This had been thought an adequate assumption for emitters with apex radii of curvature greater than about 10-20 nm, but recent suggestions[14,15] are that this limit should be significantly higher.

Although there is a strong need to move forwards with FE theory, to deal more accurately and comprehensively with data interpretation from needle-shaped and post-shaped emitters, and some initial steps have been taken (e.g., refs. 4, 15), the present paper focuses on difficulties that exist within the framework of the planar emitter approximation. My thinking is that the outcome of the present proposals, together with other[16,17] and further improvements in the use of this approximation, can then be put together to create a better background for future work.

It also needs stressing that all the FE equations discussed below, except eq. (9) as used in some contexts, are strictly applicable only to free-electron metals. However, it has long been normal practice, particularly in FE data analysis, to apply these equations to both real metals and to non-metals "as a first approximation" (often without any deep enquiry as to whether this is really justified). In this context, the aim here is to improve the quality of this "first approximation".

### B. Core FE theory equations

In any theory of FE, a key component is the "core" equation that gives the local emission current density (LECD) $J_L$ in terms of the local work function $\phi$ and the local value $F_L$ of surface electric field.

"Auxiliary" equations are also needed, in order to relate a characteristic value of local field to the measured voltage, and the measured current to a characteristic LECD value. These auxiliary equations can be written in various different forms, but are not of significant concern here. For the core FE equations, many different forms and approximations have been proposed, even for the smooth planar emitter-surface models under discussion here (e.g., see ref. 18). Four physically different core FE equations are relevant to the present discussion.

All FE equations employ in their derivations some model for the variation with distance of a quantity $M$ that describes the tunneling barrier and is called here the *electron motive energy*. In planar-geometry models, this quantity is given by $M(z) = U_e(z) - E_n$, where $U_e(z)$ describes the variation of electron potential energy with distance $z$ measured normally outwards from the emitter's electrical surface, and the normal-energy $E_n$ is the component (of the tunneling electron's total energy) associated with motion normal to the emitter surface. Both $U_e(z)$ and $E_n$ need to be measured from the same energy reference level, often taken as the emitter's Fermi level (in which case the normal-energy is denoted by $\varepsilon_n$.). The barrier is the region where $M(z) \geq 0$.

Derivations of eqns (3) and (4) below are based on the *exactly triangular (ET) barrier* given by

$$M^{ET}(z) = H - eF_L z, \quad (z \geq 0), \tag{1}$$

where $e$ is the elementary positive charge, and $H$ [$\equiv \phi - \varepsilon_n$] is the zero-field barrier height. By contrast, derivations of the Murphy-Good-type equations (5) and (9) below are based on a motive-energy expression that includes a planar image-PE term. A suitable definition is

$$M^{SN}(z) = H - eF_L z - e^2/16\pi\varepsilon_0 z, \quad (z > z_c), \tag{2a}$$

$$M^{SN}(z) = -K_F, \quad (z \leq z_c), \tag{2b}$$

where $\varepsilon_0$ is the vacuum electric permittivity, $K_F$ is the Fermi energy, and $z_c$ is a *cut-off distance* defined by applying the criterion $M^{SN}(z_c) = -K_F$ to eq. (2a). This motive-energy form is sometimes called the *Schottky-Nordheim (SN) barrier*.

The *original 1928 Fowler-Nordheim (FN) FE equation* has the form[5,19,20]

$$J_L^{orig} = P_F^{FN} a\phi^{-1} F_L^2 \exp[-b\phi^{3/2}/F_L], \tag{3}$$

where $a$ and $b$ are the FN constants[20] (see Table 1), and $P_F^{FN}$ is a field-independent pre-factor[5,20] (its exact form is of no interest here).

Table 1. Values of FE universal constants used in this paper.

| Name | Symbol | Numerical value | Units |
|---|---|---|---|
| First FN constant | $a$ | 1.541 434 | μA eV V$^{-2}$ |
| Second FN constant | $b$ | 6.830 890 | eV$^{-3/2}$ V nm$^{-1}$ |
| Schottky constant[a] | $c$ | 1.199 985 | eV (V/nm)$^{-1/2}$ |
| - | $c^2/2e$ | 0.719 9823 | eV nm |
| - | $bc^2$ | 9.836 238 | eV$^{1/2}$ |
| - | $ac^{-4}$ | 7.433 978×10$^{11}$ | A m$^{-2}$ eV$^{-3}$ |

[a]In the ISQ, the Schottky constant $c$ appears in the Schottky-effect formula $\Delta_s = cF^{1/2}$, which gives the barrier reduction $\Delta_s$ induced by field $F$.

In the *elementary FE equation*, the pre-factor is omitted, leaving the equation

$$J_L^{el} = a\phi^{-1}F_L^2 \exp[-b\phi^{3/2}/F_L] \,. \tag{4}$$

This simplified equation is convenient for undergraduate teaching, since a simplified derivation can be demonstrated without the use of research-level mathematics, but it also became widely used in the research literature after about 2005, possibly partly as a result of a poorly-worded statement in a widely cited paper relating to FE from carbon nanotubes[21]. It is this equation that is the widely used "defective equation" (perhaps better described as a "defective research equation") discussed above.

The (1956) *Murphy-Good (MG) zero-temperature FE equation*[9,22] is conveniently written in the linked form

$$J_L^{MG0} = t_F^{-2} J_{kL}^{SN} \,, \tag{5a}$$

$$J_{kL}^{SN} \equiv a\phi^{-1}F_L^2 \exp[-v_F b\phi^{3/2}/F_L] \,, \tag{5b}$$

where $J_{kL}^{SN}$ is the so-called *kernel current density for the SN barrier*, and $v_F$ is a particular value (appropriate to a SN barrier characterized by $\phi$ and $F_L$) of a special mathematical function[23] v($x$). This function v($x$) is a special solution of the Gauss Hypergeometric Differential Equation, and the *Gauss variable x* is the independent variable in this equation. $t_F$ is a particular value of a second special mathematical function $t_l(x)$, defined by eq. (7) below.

For operation at non-zero temperatures, a temperature-dependent correction factor should in principle be inserted into eqns (3), (4) and (5a). However, at room temperature this is always small (<1.2), so the factor is nearly always omitted.

In the 1956 derivation[9] of eq. (5), the parameters $v_F$ and $t_F$ were given by formulae expressed in terms of complete elliptic integrals of the first and second kinds. Subsequently, various simple two-term (and other) approximations were developed. However, mathematical developments[22-24] in the last ten years or so have uncovered: (a) the form of an exact series expansion[23] for v(x); (b) a high-precision numerical formula[22] for v(x), accurate in $0 \leq x \leq 1$ to better than $8 \times 10^{-10}$; and a simple good approximation[24] for v(x) that is accurate[22] in $0 \leq x \leq 1$ to better than 0.33 %, namely

$$v(x) \approx 1 - x + (x/6)\ln x . \qquad (6)$$

At a corresponding level of approximation, a function $t_1(x)$ (or simply t(x)) is defined by (e.g., ref. 22) and given by another simple good approximation:

$$t_1(x) \equiv v(x) - (4/3) x \, dv/dx \approx 1 + (x/9) - (x/18)\ln x . \qquad (7)$$

It has also been formally shown (e.g., see Section IVB in ref. 25, and set $\mu = f$) that these formulas are applied to MG theory by setting $x = f$, where the *scaled field* $f$ (for a SN barrier of zero-field height $\phi$) is defined by

$$f \equiv c^2 \phi^{-2} F_L \cong [1.439965 \text{ eV}^2 \text{ (V/nm)}^{-1}] \phi^{-2} F_L , \qquad (8)$$

where $c \; [\equiv (e^3/4\pi\varepsilon_0)^{1/2}]$ is the Schottky constant (see Table 1).

A weakness of both the FN and MG treatments is that they disregard the existence of atoms, by considering smooth structureless flat surfaces, rather than crystallographically structured flat surfaces. To recognize the quantitative uncertainty resulting from this approximation (and others), the present author has introduced a version of eq. (5a) in which $t_F^{-2}$ is replaced by a correction factor (or "knowledge uncertainty factor") $\lambda$ of unknown exact value (e.g., see ref. 18). This results in the so-called *Extended Murphy-Good (EMG) FE Equation*[16]

$$J_L^{\text{EMG}} \equiv \lambda a \phi^{-1} F_L^2 \exp[-v_F b \phi^{3/2}/F_L] . \qquad (9)$$

Current thinking[26], in 2019, is that $\lambda$ probably lies in the range $0.005 < \lambda < 14$, though the lower limit may eventually prove to be unnecessarily low.

It can be seen that, in both eq. (5) and eq. (9), an effect of including the image-PE term in the motive energy has been to insert a mathematical correction factor $v_F$ into the exponent of the FE equation. As shown next, it is this factor $v_F$ that is primarily responsible for increasing the current density by an *improvement factor K* roughly of order 300 .

## III. ANALYSIS

## A. Derivation of improvement factor values

Another feature of the mathematical developments in FE theory in the last ten years has been the realization that a very useful *scaled form*[27] exists for the kernel current density $J_{kL}^{SN}$. This is obtained by rewriting eq. (8) as $F_L = c^{-2}\phi^2 f$, inserting this expression into eq. (5b), and defining work-function-dependent scaling parameters $\theta$ and $\eta$ by (see Table 1)

$$\theta = ac^{-4}\phi^3, \quad \eta = bc^2\phi^{-1/2}. \tag{9}$$

This results in the scaled equation

$$J_{kL}^{SN} = \theta f^2 \exp[-v(f)\cdot \eta/f], \tag{10}$$

where $v_F$ has here been written explicitly as a function $v(f)$ of $f$. A work-function of 4.500 eV yields the values $\eta \cong 4.3868$, $\theta \cong 6.774 \times 10^{13}$ A/m$^2$. This scaled form, which involves only a single independent variable, is often useful for algebraic manipulations and (as here) for simple numerical calculations. Part of its usefulness is that, in experiments on "ideal emitters"[16], the scaled field $f$ can be *measured* reasonably well by using a Fowler-Nordheim plot (or a Murphy-Good plot[16]), and that $f$ has an obvious and easy interpretation ("to get the corresponding local barrier field, multiply $f$ by the reference field $F_R [=\phi^2/c^2]$"). When $\phi$=4.50 eV, then $F_R \approx 14.1$ V/nm.

For the exactly triangular barrier, the corresponding scaled equation is obtained by replacing $v(f)$ in eq. (10) by unity, to give

$$J_L^{el} = \theta f^2 \exp[-\eta/f]. \tag{11}$$

Setting $x=f$ in eq. (6) yields an expression for $v(f)$. Substitution into eq. (10), and some trivial algebraic re-arrangement, yields

$$J_{kL}^{SN} \approx \theta f^2 f^{-\eta/6} \exp\eta \, \exp[-\eta/f]. \tag{12}$$

The prefactor $t_F^{-2}$ will make a very small contribution, but by far the largest contribution to the improvement factor $K$ comes from the ratio $J_{kL}^{SN}/J_L^{el}$, which is given approximately by

$$K = J_{kL}^{SN} / J_L^{el} \approx f^{-\eta/6} \exp\eta. \tag{13}$$

This expression provides qualitative understanding of the origin of the improvement factors. For $\phi$=4.500 eV, the factor $\exp\eta$ has a value near 100; the other factor provides some additional improvement.

However, better numerics are obtained by using the high-precision formula[22] for v(*f*) and a related formula for *K*, namely

$$K = J_{kL}^{SN}/J_L^{el} = \exp[\{1-v(f)\}\cdot\eta/f]. \tag{14}$$

Some illustrative values are shown in Table 2. They have been calculated by coding the high-precision formula for v(*f*) [=v(*x*=*f*)] into an Excel™ spreadsheet, which is provided as supplementary material. The *f*-values used are based on the knowledge[27] that, for tungsten, operating *f*-values are often within the range 0.15<*f*<0.35, corresponding (with *ϕ*=4.50 eV) to a barrier-field range 2.1 V/nm < $F_L$ < 4.9 V/nm. It can be seen that a "typical" value for *K* is around 300, (though *K* actually varies from around 250 to around 500 for the *ϕ*-value and range of *f*-values chosen).

Table 2. To show values of v(*f*) and the improvement factor *K*, for values of the scaled field *f* and corresponding local field $F_L$.

| *f* | $F_L$ (V/nm) | $v_F$=v(*f*) | *K* |
|---|---|---|---|
| 0.15 | 2.1 | 0.800 | 481 |
| 0.20 | 2.8 | 0.744 | 378 |
| 0.25 | 3.5 | 0.690 | 314 |
| 0.30 | 4.2 | 0.638 | 270 |
| 0.35 | 4.9 | 0.587 | 238 |

## B. Validity of using classical image potential energies

As indicated above, the correction factor $v_F$ in eqns (5) and (9) derives from the inclusion of the classical image-PE term ($-e^2/16\pi\varepsilon_0 z$) in the motive energy. Although originally discussed as an obvious classical correction, it has long been understood that this term is a model for the quantum-mechanical (QM) exchange-and-correlation (E&C) effects experienced by electrons in real metals. A priori, it is not obvious that the classical image PE would necessarily be a good model for E&C effects at surfaces. The issue is whether the classical image PE term is a "sufficiently good" approximation, or whether it would be better to leave it out.

The underlying physics was first addressed in detail in a paper[28] written by Bardeen in 1940. [Bardeen is the only person to have been awarded the Nobel prize in Physics twice. One of these (with Shockley and Brattain) was for the practical realization of the transistor, where a significant contribution from Bardeen was to develop understanding of how electron surface states affected transistor operation. Kroemer, one the three scientists who found the mathematical mistake in the Nordheim 1928 paper, also (much later) received a Nobel prize for work on transistor development.]

Put simply, Bardeen's relevant conclusion in his 1940 paper was that, as distance outwards from a smooth metal surface increases, the QM correlation contribution (to the total electron PE that needs to be inserted into the relevant one-electron Schrödinger equation) tends to become equal to the classical image PE. Much later (see Chap. 12 in ref. 29, and references therein), it was established that the QM exchange contribution also tends to become equal to the classical image PE at large distances. The issue thus becomes: where is the FE tunneling barrier, relative the emitter surface atoms?

In the context of the planar emitter approximation, this issue divides into two separate questions: (a) Where is the emitter's electrical surface (i.e., the reference surface from which the *electrical distance z* is measured) relative to the surface-atom nuclei? (b) What are the electrical distances for the inner ($z_{in}$) and outer ($z_{out}$) edges of the tunnelling barrier?

For a SN barrier of zero-field height $\phi$, there are well established expressions for $z_{out}$ and $z_{in}$ ($z_{out}$>$z_{in}$). These can be written (e.g., see eq. (45) in ref. 25, and put $H=\phi$, and $\mu=f$):

$$z_{out}, z_{in} = (\phi/2eF_L) [1\pm(1-f)^{1/2}] , \qquad (14)$$

On substituting $F_L = c^{-2}\phi^2 f$, this becomes

$$z_{out}, z_{in} = (c^2/2e\phi) [(1/f \pm (1/f^2 - 1/f)^{1/2}] \cong (0.7199823 \text{ nm})\cdot(\text{eV}/\phi)\cdot[(1/f \pm (1/f^2 - 1/f)^{1/2}] , \qquad (15)$$

which for $\phi = 4.500$ eV reduces to

$$z_{out}, z_{in} \cong (160.0 \text{ pm})\cdot[(1/f \pm (1/f^2 - 1/f)^{1/2}] . \qquad (16)$$

Resulting values are shown in Table 3. For comparison purposes these are shown to a precision of 1 pm, but this is not a statement of physical accuracy (which is difficult to assess reliably).

With planar surfaces, the *electrical surface* is the "reference plane" that $z$ must be measured from if the electrostatic potential ($-eF_L z$) is to have its correct value at large distances. In the simplest models, image effects are also measured from this surface[29,30]. There is a "repulsion effect" that moves the electrical surface outwards from the plane of the surface-atom nuclei, by a distance $d_{rep}$ called the *repulsion distance*[31-33].

Surprisingly, the issue of electrical-surface location has rarely been addressed in the specific context of field electron emission, though there are two recent discussions[33,34]. However, it is a well known issue in field ion emission, where it is known that its location coincides with the centroid of the induced charge[30,35]. Estimates have been made of $d_{rep}$ for atomically structured planar surfaces, using relevant crystallographic information about surface structure. The simple classical model originally used[31,35], which attributes the repulsion effect to the electric dipole moment generated by the field-induced electrostatic polarization of the metal surface atoms (including mutual depolarization effects), is applicable to both positively and negatively charged surfaces, and for the tungsten (110) surface yields the value 157 pm.

Table 3. To show values of parameters related to the locations of the inner and outer edges of a SN tunneling barrier of zero-field height 4.50 eV, with $d_{rep}$=157 pm. [See text for explanation of notation.]

| $f$ | $z_{in}$ (pm) | $z_{out}$ (pm) | $d_{in}$ (pm) | $d_{out}$ (pm) |
|---|---|---|---|---|
| 0.15 | 83 | 2050 | 240 | 2207 |
| 0.20 | 84 | 1516 | 241 | 1673 |
| 0.25 | 86 | 1194 | 243 | 1351 |
| 0.30 | 87 | 980 | 244 | 1137 |
| 0.35 | 89 | 826 | 246 | 983 |

Recently[34], density functional theory (DFT) calculations on a tungsten double pyramid have yielded estimates of 161 pm for the repulsion distance above the top atom on the positively charged side, and 213 pm for the repulsion distance above the top atom on the negatively charged side. (The difference occurs because, due to exchange repulsion forces—which obviously are not included in the classical model—it is easier for an impressed electrostatic field to "pull electrons out" than "push electrons in".)

For safety, we take $d_{rep}$ as equal to the lower of the estimates for a negatively charged surface, namely 157 pm. We can then make estimates of the distances (from the plane of the surface-atom nuclei) of the inner edge ($d_{in}$) and outer edge ($d_{out}$) of the tunneling barrier. These estimates are shown in Table 3, and need to be compared with a distance $d_{NN}/2$ that is half the nearest-neighbor separation $d_{NN}$ in the tungsten lattice, and is a measure of "effective atomic radius".

For tungsten, $d_{NN}/2$= 137 pm. Table 3 thus shows that, in all practical operating cases, the inner edge of the tunneling barrier is "right at the edge of the electron charge-clouds", and that most of the barrier region is "well away from the atom nucleus". The usual FE theoretician's conclusion, albeit a qualitative one, is as follows. All of the barrier region is sufficiently far away from the dense electron charge-clouds around the nearest atomic nuclei that the approximation that "the PE variations predicted by QM E&C effects are close to those predicted classically" is likely to be reasonably good, and hence that eq. (2a) is a good approximation. Expectation (from use of eq. (15) and the classical model) is that all metals should exhibit broadly similar behavior,

Historically, there have been a few dissenting opinions, in particular those of Hartstein and Weinberg[36] and of Ancona[37], but these have not found significant acceptance. It can be validly argued that issues of detail remain to be fully resolved, and that a slightly modified barrier (particularly on the inner side) might be an even better choice than the SN barrier. However, these things do not affect the conclusion that, in the choice between the two widely used barriers (ET and SN), it seems significantly better physics to choose the SN barrier, and hence choose Murphy-Good-type equations (with the factor $v_F$ in the exponent).

## IV. DISCUSSION

### A. The difficulties of experimental verification

In the ideal world, the above issue would be settled by experiment, but in fact it is very difficult to do this. At a casual level, if some experimental effect that depends on current density can be explained by eq. (5), then it can also be explained by eq. (4) if you take the field to be somewhat higher and/or the emission area to be somewhat larger. So one then needs to have accurate *independent* methods of measuring local electrostatic field and emission area. But independent methods have not yet been applied, either because they do not exist, or because they are not yet sufficiently well developed with *demonstrably* sufficient accuracy.

Another complication arises when eq. (9) is used: it is possible to imagine factors (such as "poorer wave-matching at the surface when atomic-level wave-functions are used") that would cause $\lambda$ to be less than unity: this would tend to counterbalance effects due to the improvement factor $K$, and thus contribute to the difficulties of experimental verification.

An alternative approach would be to plan experiments specifically to test the theory. But here the need would be to compare experimental results with deductions from both alternative theories, and to show that theory based on the SN barrier is significantly superior. For example, the field electron energy distribution measurements of Gadzuk and Plummer[38], when compared with predictions from MG theory, are sometimes offered as evidence in support of MG theory. They do provide evidence of this. But reality is that a theory of tunneling though an ET barrier leads to generally similar results, and comparisons as currently made are not good enough to distinguish decisively between the two barrier models[39].

A wider (but incomplete) survey of possible methods was undertaken some years ago[39], with eight possible methods identified. Overall, the evidence seemed mildly in favour of the SN barrier and MG theory, but it was difficult to conclude, for any individual method, that currently available experimental evidence and analysis led to a decisive outcome.

This issue of how to establish a decisive *experimental* test of the proposition that the Murphy-Good equations are superior to the original Fower-Nordheim ones is "unfinished business". It may be easier to formulate a decisive test of this type when issues related to data-analysis from needle-shaped and/or post-shaped emitters are better understood, with one possible route (Popov, Kolosko, Filippov, and Forbes, unpublished work) being measurement of the value of $k$ in the empirical FE equation for measured current-voltage $I_m(V_m)$ data, namely

$$I_m = CV_m^k \exp[-B/V_m] \, , \qquad (17)$$

where $B$ and $C$ are initially treated as constants. The main present difficulty with this approach seems to be getting experimental $I_m(V_m)$ data that is sufficiently noise free.

### B. The views of neighbouring communities

For the ET barrier to be better physics than the SN barrier, something like the following proposition would need to be true. "That, in a quantum-mechanical world the classical Schottky effect is not a real effect but an artefact of classical thinking, and that all phenomena attributed to the Schottky effect are in fact due to tunneling through an exactly triangular barrier."

There are several contexts in physics, in particular in electronic device theory, where it is assumed that the Schottky effect operates, for example the theory of photocathodes and the theory[40,41] of the commercial electron microscope electron source type known as the "Schottky emitter". In the latter case, it is assumed that electrons escape both over and through a SN barrier, and the theory of the beam energy spread (which affects the focussing properties of the electron beam) is developed accordingly[40,41]. Both these applications are the basis for highly successful commercial technologies. In such contexts, the proposition that the Schottky effect is an artefact of classical thinking would, almost certainly, receive little or no credibility.

### C. Intermediate conclusions

In summary, the current position appears to be as follows.

(1) From the theoretical point of view, there is very strong evidence that the SN barrier is better physics than the ET barrier, and hence that the Murphy-Good-type FE equations are better physics than the ET-barrier-based equations.

(2) At present, it is difficult to conclude that there is experimental evidence decisively in favor of the SN barrier as being better physics, but against this there is no known experimental evidence that is decisively in favor of the ET barrier.

(3) There are successful neighboring communities that design, build and sell devices with theory that assumes that the Schottky effect is real and (in the case of Schottky emitters) that tunneling through a SN barrier occurs.

In these circumstances, it seems fairly clear that the better scientific position, at present, is to accept that the SN barrier is better physics than the ET barrier. In any case, to do otherwise would disturb a lot of physics, going back to the work of Schottky[42,43], and before him to Kelvin[44], Maxwell[45] and J.J. Thomson[46].

As things stand at present, it is perhaps not totally impossible in principle that a valid scientific case could be made for preferring the ET barrier over the SN barrier, at least in some circumstances, but the onus is on those who believe that "Fowler-Nordheim is better than Murphy-Good" to make a case that is compatible with all the existing evidence.

Note that this discussion has been about emission from metals. There is some evidence[47] that exchange-and-correlation effects outside a graphene sheet or a carbon nanotube cannot be well represented by a classical image PE. But even in this case the SN barrier is probably a better starting point than an ET barrier, with the most obvious simple model for such a situation being one in which a "weakening factor" $w$ is included in the image PE, which then becomes $(-we^2/16\pi\varepsilon_0 x)$.

### D. Issues relating to nomenclature and citation practice

The author is inclined to think is that the primary cause of the pathological situation discussed in this paper is often lack of awareness. Two practices of the wider FE and electron emission communities may contribute significantly to this. The first is the system for naming equations, which tends to be "ancestor-oriented", both in field electron and in thermal electron emission. The various equations that the present paper refers to as FE equations are often all called "the Fowler-Nordheim Equation". The problem is that different groups of researchers tend to apply this name to physically different equations: most FE theoreticians regard eq. (5) as "the FN equation"; many FE papers concerned with technological emitter development regard eq. (4) as "the FN equation"; and non-experts sometimes regard eq. (3) as "the FN equation" (which would probably be closest to normal scientific practice outside the FE community). Other authors give the name "the FN equation" to approximated or mathematically defective versions of eq. (5).

Apart from generating confusion, this nomenclature practice tends to direct attention back towards the original 1928 FN paper, with its known weakness in physical thinking, and to disguise the fact that there are numerical differences between the predictions of the various equations. The author's present view is that all these (and the many other equations used to describe FE) are better described as "field emission equations" or "field electron emission equations" (with both terms abbreviated to "FE equations"). The different core equations should be given distinctive names, incorporating the names of the relevant authors where helpful. If it is wished to indicate a relationship to the seminal 1928 work of Fowler and Nordheim, then some phraseology similar to the following could be used:   "… is based on the Murphy-Good zero-temperature FE equation [**], which corrected poor assumptions made in the physical thinking behind the derivation of the original 1928 Fowler-Nordheim FE equation [**], …" This kind of approach would make the true scientific situation much clearer.

A second unhelpful feature, particularly in some FE technological literature, is the practice of giving a reference to the original 1928 FN paper as the only theoretical reference in a technological article, with no reference given to any subsequent theoretical development carried out in the last 90 years, or to any FE research textbook. (For example, see ref. 48, but there are many other examples.) It is unfortunate that modern reviewing systems regularly allow this. A major problem is that the weakness in the 1928 paper is as follows (see Appendix A): FN clearly state that corrections of the type subsequently considered by Murphy and Good are small and not important: obviously, this gives the impression that use of the exactly triangular barrier is an adequate approximation. As shown earlier in this paper, this is not in fact true in modern contexts (unless you happen to think that discrepancies by a factor of up to around 500 are not important). By failing to give the references that would allow other scientists to judge for themselves, the many (though certainly not all) FE technological papers that do this—and the reviewing systems that facilitate it—are accidentally allowing modern FE scientific practice to become determined by "1928 thinking" rather than by the normal scientific processes of error detection and removal.

**E.  Relevance to semiconductor literature**

Lack of awareness of the influence of exchange-and-correlation effects (as modeled by a classical image PE) on tunneling theory may also occur in other scientific and technological areas, an obvious example being the theory of nanoscale electronic devices. As integrated-circuit processing line-widths reduce towards 3 nm, it becomes obviously necessary to pay closer attention to tunneling-based mechanisms that might influence device quality or device failure rates. It would seem helpful, therefore, for undergraduate and graduate students to be made aware that image-PE-based theories might predict effects of this kind to be more severe than theory that neglects the role of the image PE.

To take one example, the well-known semiconductor-device student textbook authored by Sze and Kwok[49] disregards effects due to image PE when discussing tunneling theory. In particular, it might be helpful to improve the "tunneling" entry in Table 2 on p. 227 (where image-PE effects are included for "thermionic" emission, but not for tunneling), and to look again at equations such as eq. (41) in Section 8.3.2, in order to assess whether the inclusion of image-PE effects would have any significant consequences.

**F. Final comments**

Even though the theoretical evidence is clear and strong, my conclusion above was that, as yet (in FE), there is no decisive experimental evidence that the SN barrier is better physics that the ET barrier. There may be some who feel that it is better to stay with 90-year-old theory (rather than change to 60-year-old theory or 21st century theory) until the matter is resolved experimentally. This approach would probably be a mistake. In the planar emitter approximation (which is what most people currently use), technology development is best served by using the best theory currently available in that approximation, which means either eq. (5) or (better) eq. (9). If, when we are eventually able to interpret measured FE current-voltage data more effectively, it turns out that in fact the performance of eq. (5) is worse than that of either eq.(3) or eq. (4), then the best interpretation will almost certainly be, not that the SN barrier is worse physics that the ET barrier, but that some other factor (known or unknown) is influencing the value of $\lambda$ in eq. (9). To stay with eq. (3) or eq. (4) at the present time would likely make it more difficult to improve FE theory further in the longer term.

For the avoidance of doubt, it is necessary to stress that there is no suggestion whatsoever here of any scientific dishonesty—rather, the issue is one of collective scientific misunderstandings that have given rise to pathological science (including pathological reviewing). Obviously, there is no technological reward in the use of ET-barrier-based equations rather then SN-barrier based equations. In the context of electrical breakdown, which tends to occur when current exceeds a certain level, there is no merit in writing down an equation that under-predicts current levels by a factor of 300 or thereabouts. Likewise, when attempting to develop/investigate new emitter materials for high-current electron sources, there is no merit in writing down an equation that under-predicts the performance of the material/device by a factor of 300 or thereabouts.

In the absolute scale of things, this error by a factor of 300 or thereabouts may not be that important, but it is an unnecessary and avoidable error that also tends to discredit the process of scientific reviewing. In principle, this error could easily be taken out of future FE literature by improved knowledge of FE theory amongst authors and by enhanced care in reviewing, making use of the detailed discussion given in Section 3. I would encourage my FE colleagues, when acting as reviewers, to assist in the process.

**APPENDIX A: WEAKNESSES IN THE 1928 PAPERS**

Although other discrepancies exist in the original 1928 Fowler-Nordheim paper[5], only those relating to barrier shape are of interest here. Fowler and Nordheim were well aware that image-type effects would round the tunneling barrier apex. But Fowler must also have known from previous experience[50] and contact with Jeffreys (both were members of Trinity College, Cambridge) that, when the image PE term is included in the one-dimensional Schrödinger equation, then the resulting second-order differential equation had no known analytical solutions in terms of the established functions of mathematical physics. (This is still the case 90 years later.)

However, the Schrödinger equation form derived from the ET barrier does have exact solutions, and is thus an obvious first case to be analyzed. The issue is whether image rounding of the barrier apex would have any significant influence. After brief earlier comments, FN return to this in their "§4 *Additional Calculations*". They use the following notations. Their symbol $W$ denotes normal kinetic energy, i.e. normal-energy $E_n$ measured relative to the base of their PE well, their symbol $C$ denotes inner potential energy, as shown in Fig. (1a) below, their $\kappa$ denotes the "Schrödinger equation constant" $(2m)^{1/2}/\hbar$, where $m$ is electron mass and $\hbar$ is Planck's constant divided by $2\pi$, their $Q$ denotes a parameter equal to the modern $(bH^{3/2}/2F)$, their $D$ denotes tunneling probability, and their "†" relates to a (1925) paper by Jeffreys[51] that describes the mathematical approximation later known as the "JWKB" (or "WKB") approximation. They write the following.

"It is not difficult to show generally, by comparisons with Jeffreys'† asymptotic solution of a similar equation, that for values of $W$ not too near C the exact form of the potential peak and the rounding off at the top will not seriously affect the emission coefficient $D(W)$. Our direct calculation for this very simple case is therefore sufficiently typical. For the $Q$ of an exact solution has to be replaced by

$$\kappa \int_{x_0}^{x_1} (V-W)^{1/2} dx \ ,$$

where $V$ is the potential energy of the electron at any point and $x_0$ and $x_1$ are the points at which $V$–$W$ vanishes. The integration range is shown in fig. 3 [our Fig. 1a, below]. It is at once clear that, provided the shaded area is reasonably large, modifications in the contour near the peak are unimportant."

Unfortunately, FN seem to have grossly underestimated the amount of rounding that occurs. Nowadays, it is easily shown by spreadsheet calculations that, for work function 4.50 eV and a barrier field of 5 V/nm (about the worst case for the typical operating range), the true shape of a SN barrier is as shown in Fig. 1b. The discrepancy is obvious, the barrier is in fact much weaker than FN assumed, and the "contour modifications" are in fact important.

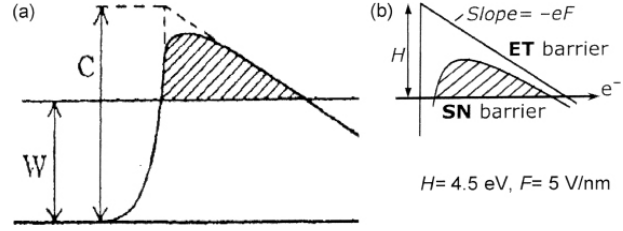

FIG. I. Comparison of Schottky-Nordheim barriers: (a) as drawn by Fowler and Nordheim[5] in their Fig. 2(ii); (b) as calculated from eq. (2), using $H$= 4.5 eV, $F_L$= 5 V/nm (normally the worst-case discrepancy). The horizontal axis denotes distance and the vertical axis normal-energy. Scales are adjusted so that $H$ and $F_L$ are similar for both diagrams. There is an obvious discrepancy in the (hatched) "area under the barrier", as between the two diagrams. This converts into a strong difference in tunnelling probability.

As indicated earlier, a related paper by Nordheim[7], emerged later in 1928. This attempted to apply the theory of Jeffreys[51] (the modern first-order, or "simple", JWKB approximation) to the SN barrier. The SN barrier shape drawn in this second paper is realistic (see his Fig. 5). The first part of his derivation, from eq. (47) to eq. (52), although expressed in an older equation system, agrees with modern analysis. But, unfortunately, a mathematical mistake entered at eq. (53), where an expression given for the elliptic modulus $k$ is in fact an expression for the elliptic parameter $m$ $[\equiv k^2]$.

This mistake is easier to see when it re-appears in eq. (55), but a notation change is helpful. The symbol $x$, used by Nordheim for what was later called the *Nordheim parameter*, is here replaced by the modern notation "$y$". In the modern "International System of Quantities" (ISQ) (based on including $\varepsilon_0$ in Coulomb's Law), $y$ is given by

$$y \equiv (e^3 F_L/4\pi\varepsilon_0 \phi^2)^{1/2}. \tag{A1}$$

[This means that, numerically, $y$ is related to the scaled field $f$ by $y=+f^{1/2}$ or to the Gauss variable $x$ by $y=+x^{1/2}$].

The relevant part of Nordheim's eq. (55) can then be written:

$$k = 2(1-y^2)^{1/2}/[1+(1-y^2)^{1/2}]. \tag{A2}$$

This formula remained in the literature for 25 years, until Burgess, Kroemer and Houston[8] (BKH) pointed out, in 1953, that eq. (A2) is not correct, and should be replaced by

$$k^2 = 2(1-y^2)^{1/2}/[1+(1-y^2)^{1/2}]. \tag{A3}$$

In passing, note that in eq. (A2), and in all other relevant equations in Nordheim's paper, the Nordheim parameter appears only as its square. Thus, Nordheim could alternatively have chosen to formulate the theory in terms of the parameter $x = y^2$. With hindsight, his choice of the parameter $y$ was unfortunate, because we now know that the FE special mathematical function v($x$) (where $x$ is the Gauss variable) is a special solution of the Gauss Hypergeometric Differential Equation HDE): it is *mathematically perverse* to express this solution as a function of the square root of the independent variable in the Gauss HDE.

Some consequences of the Nordheim calculations need discussion. My general notation for the *physical/modeling* correction factor that appears in the exponent of FE equations based on the SN barrier is $v_F^{SN}$. As Table 4 shows, the above error made the value of $v_F^{SN}$ closer to unity than the correct (BKH) theory does. This has a corresponding effect on predicted improvement-factor values, and hence on predicted LECDs. (Slightly different values would be derived for the N28 quantities if values tabulated later by Houston[52] were used.)

| | | | | | |
|---|---|---|---|---|---|
| Table 4: Comparison of exponent correction factor ($v_F^{SN}$) and improvement factor ($K$) values predicted by Nordheim (1928) theory and by BKH (1956) theory. | | | | | |
| $y$ | $f$ | N28 $v_F^{SN}$ | N28 $K$ | BKH $v_F^{SN}$ | BKH $K$ |
| 0.40 | 0.16 | 0.849 | 80 | 0.7888 | 456 |
| 0.50 | 0.25 | 0.781 | 58 | 0.6900 | 314 |
| 0.60 | 0.36 | 0.696 | 50 | 0.5768 | 233 |

Notwithstanding that Nordheim's treatment is mathematically incorrect, it does generate improvement factors of significant size. But Fowler, in the second (1936) edition[53] of his textbook on Statistical Mechanics, although citing Nordheim's 1928 work, reproduced a diagram similar to that in the original 1928 FN paper (which seems to be an oversight), and continued to state that, although there is a small correction to be made by including the image effect, this was not important.

This suggests that Fowler's criterion of "importance" may have differed from that applicable to modern contexts. This is understandable. Relevant aspects of the metal-physics/emission-physics situation immediately before the FN work were as follows. (1) The physical origin of what was then called "auto-electronic emission" (modern FE) was not understood. (2) Lauritsen[54] had just found experimentally that, when auto-electronic-emission current-voltage data were plotted in the form ln{$I_m$} vs $1/V_m$, then apparently straight-line plots were generated. (3) It was thought that thermal electron emission was essentially different from the auto-electronic emission of conduction electrons, and was due to the emission of "thermions", i.e. electrons in some form of specially excited state within the metal, different from that of the conduction electrons (for example, see ref. 54, and remember that Millikan had received a Nobel prize in 1923 for experimentally verifying Einstein's predictions about the photoelectric effect, and for measuring the electron charge).

Historically, the FN paper had three main achievements. It showed that auto-electronic emission was in fact a wave-mechanical tunneling process (but credit for this should be partially shared with Oppenheimer[55-57]). This explained the Lauritsen straight-line plots. It explained (on the basis of Fermi-Dirac statistics) why auto-electronic emission currents were nearly temperature independent. And, probably most important historically, the FN paper argued (albeit briefly) that there was no need to postulate the existence of thermions: rather, both auto-electronic emission and the emission of thermions could be explained as the emission of conduction electrons from (what would now be called) a *single* electron energy band, under different conditions of field and temperature.

Thus, the FN paper, together with Fowler's earlier paper[58] introducing electron spin into metal physics and emission physics, and with the related work of Fermi[59], Dirac[60], Sommerfeld[61,62] and Bethe[62], were between them the foundational papers for modern metal-band theory and large parts of modern condensed-matter physics.

It is understandable that, against the background of this major achievement, Fowler might—with considerable justification—have regarded temporary inaccuracies in calculating FE local current densities as not of any great importance, and be keen to stress this. It is also understandable that, in the modern context of predicting electrical breakdown and electron-source performance, the removal of unnecessary errors of order 300 does seem important.

Given that modern FE literature tends to keep referring people back to the original 1928 FN paper, these historical differences may be a part cause of today's pathological FE literature. The criteria of scientific importance for FE in the 21st Century are clearly not the same as they were in the "golden years" of the late 1920s and early 1930s.